\documentclass[seceq]{ptptex}
\usepackage{epsfig}

\title{
The CGC and the Glasma:  Two Lectures at the Yukawa Insitute%
}

\author{
Larry \textsc{McLerran}%
}

\inst{
Physics Department and Riken Brookhaven Center, Brookhaven National Laboratory, Upton, NY 11973
}

\abst{
These lectures concern the theory of the Color Glass Condensate (CGC) and the Glasma.  These are forms of matter that control the earliest times in hadronic collisions.  I will motivate the CGC and Glasma from simple physical considerations, and provide a sketchy derivation from QCD.
There will be some discussion of experimental tests of these ideas.
}

\begin{document}

\maketitle

\section{Introduction}

These lectures present the theory of the Color Glass Condensate (CGC) and the Glasma in an elementary and intuitive manner. This matter controls the high energy limit of QCD.  The CGC is the universal limit
for the components of a hadron wavefunction important for high energy scattering processes.  It is a highly coherent, extremely high energy density ensemble of gluon states.  The Glasma is matter produced in the collision of CGCs of two hadrons.  It has properties much different from those of the CGC, and is produced in a very short time after the collision.  It eventually evolves from the the Color Glass Condensate initial conditions into a Quark Gluon Plasma.  
\begin{figure}[!htb]
\begin{center}
  \mbox{{\epsfig{figure=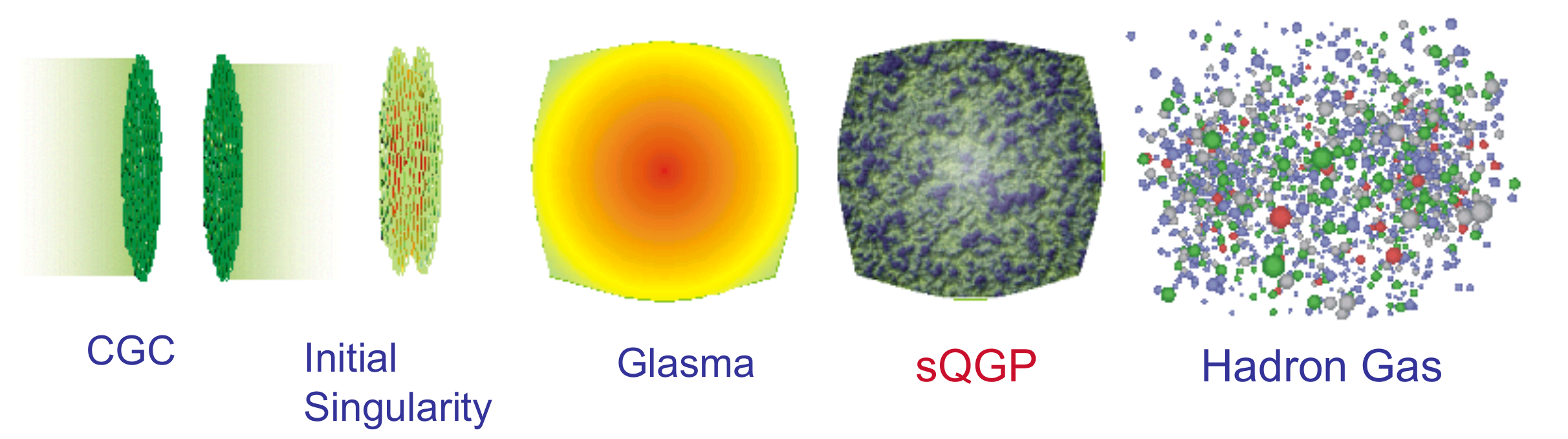, width=0.95\textwidth}}}
   \end{center}
\caption[*]{ \small  A visualization of the collision of two high energy hadrons.}
     \label{collision}
\end{figure}

We can visualize the collision of two high energy hadrons as shown in Fig. \ref{collision}.  Before the collision, two hadrons appear as Lorentz contracted sheets approaching one another at near light speed.
These we will later describe as two sheets of Colored Glass.  In a very short time, the sheets of Color Glass interpenetrate one another.  This we think of as the initial singularity for the collision.  This is of course not a real singularity for finite collision energy, but we will see it becomes one in the limit of infinite energy.
After the initial singularity, a Glasma is formed.  This is composed of highly coherent gluon fields of very high energy density.  If we imagine that the sheets of Colored Glass have passed through one another
largely intact, the Glasma forms in the region between the receding sheets.  As time goes on, the Glasma evolves into a Quark Gluon Plasma, and eventually into a gas of ordinary hadrons.

These lectures are about the earliest stages of these collisions, and will describe neither the Quark Gluon  Plasma nor the Hadron Gas.

\section{The Hadron Wavefunction at High Energy}

 \begin{figure}[!htb]
\begin{center}
  \mbox{{\epsfig{figure=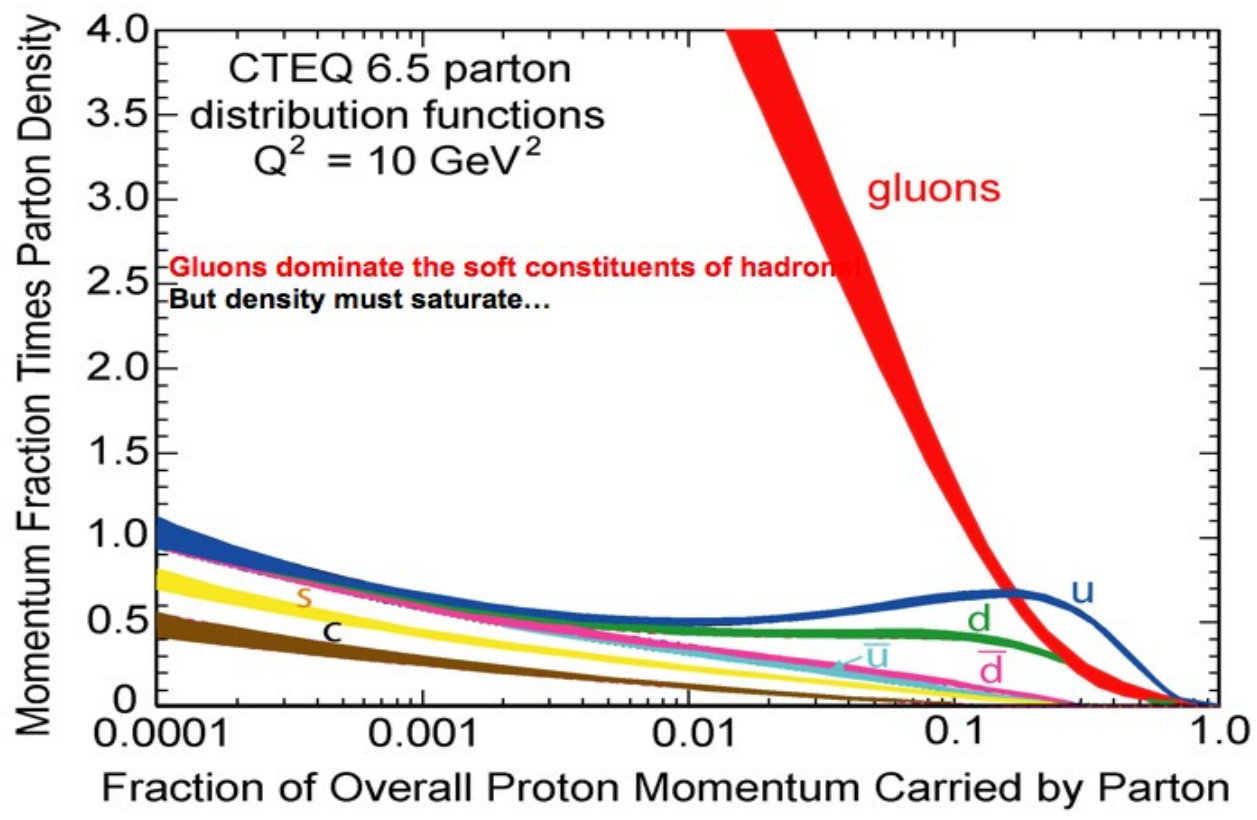, width=0.70\textwidth}}}
   \end{center}
\caption[*]{ \small  A gluons and quark composition of a  hadron as a function of $x$.}
     \label{gluons}
\end{figure}
We will imagine the hadron wavefunction decomposed in its hadronic constituents.  We will measure the number of gluons and quarks in terms of their fractional momentum, $x$ of a high energy hadron. 
\begin{equation}
    x = E_{constituent}/E_{hadron}
\end{equation} The gluon and quark distributions are shown in Fig. \ref{gluons}.  For $x \le 10^{-1}$ gluons dominate the wavefunction.
The gluon component grows without bound in as $x \rightarrow 0$.  The small x limit is also the high energy limit since for a minimal gluon energy of order the QCD scale $E_{gluon} \sim \Lambda_{QCD}$,
$x \sim \Lambda_{QCD}/E_{hadron}$.  
The size of a hadron does not grow rapidly as energy approaches infinity.  The growth of the cross
size is limited by he Forissart bound $\sigma \le \sigma_0 \ln^2(E/E_0)$.  Since the number of gluons is growing rapidly but the size of a hadron is not, the high energy limit of QCD is also the high gluon density limit.

It is important that we understand we are talking about the part of a hadron wavefunction that generates the matrix elements for highly inelastic processes.  If we were to look at processes that produced very few particles, these matrix elements might not be generated by the high gluon density component of the wavefunction.  A hadron has many Fock space components.  Those components with few gluons and
quark-antiquark pairs dominate low energy processes, and some processes not involving multiparticle production at high energy.  The components in which we are interested  dominate the small x gluon distribution function,
and more generally, highly inelastic processes.

\section{Space Time Picture of Hadronic Collisions}

Before developing the theory of the Color Glass Condensate, let us first review Bjorken's space time picture of high energy strong interactions\cite{Bjorken:1976mk}.
\begin{figure}[htb]
\begin{center}
  \mbox{{\epsfig{figure=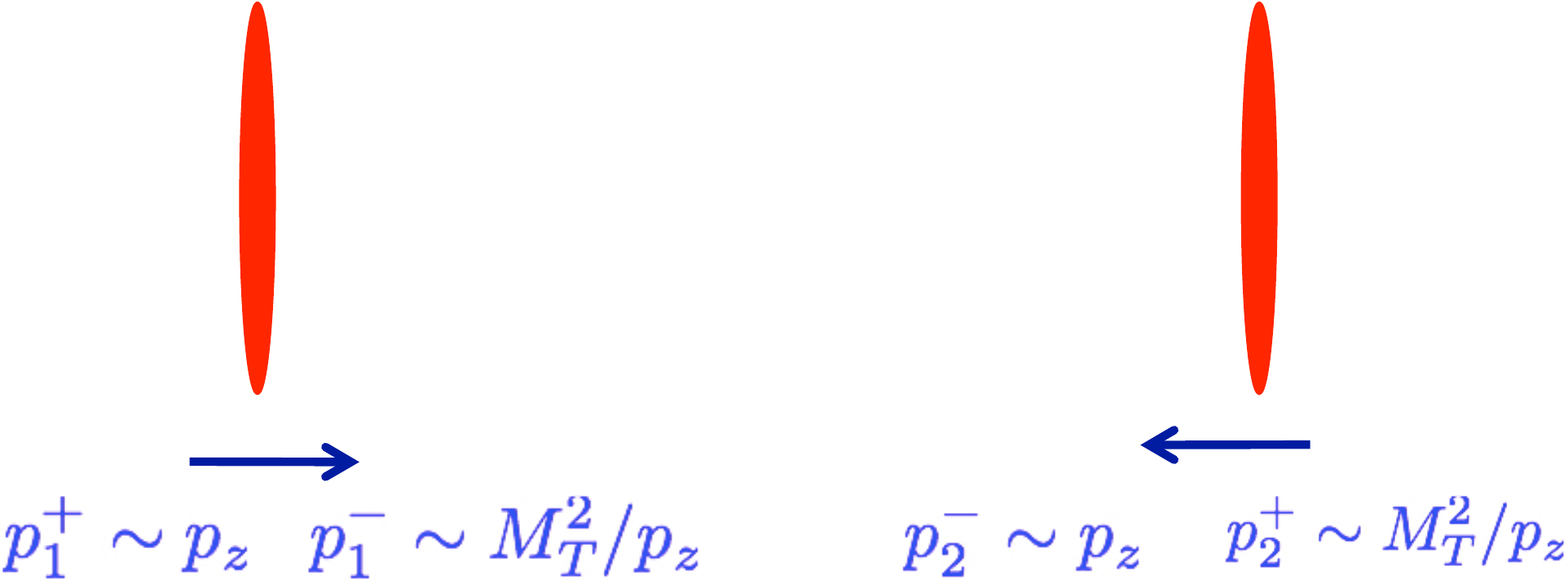, width=0.70\textwidth}}}
   \end{center}
\caption[*]{ \small  Bjorken's space time picture of high energy hadronic collisions.}
     \label{spacetime}
\end{figure}
Bjorken imagined that high energy hadrons would be Lorentz contracted sheets of partonic constituents.
Each partonic constituent is Lorentz contracted by its gamma factor,
\begin{equation}
     \gamma = x E_{hadron}/M_T
\end{equation}
where $M_T = \sqrt{p_T^2+M^2}$.
Here coordinates are measured parallel to the collision axis (z direction) and transverse to it $T$.
The description of the hadronic constituents is simplified using light cone coordinates,
\begin{equation}
      p^\pm = {1 \over \sqrt{2}} (E \pm p_z)
\end{equation}
and
\begin{equation}
  x^\pm = {1 \over \sqrt{2}} (t \pm z)
\end{equation}      
In light cone coordinates $p^+p^- = M_T^2/2$, or $2p^+p^- -p_T^2 = M^2$ and
\begin{equation}
  p \cdot x = p^+ x^- + p^- x^+  - p_T \cdot x_T
\end{equation}  
The uncertainty principle relation is
\begin{equation}
   \Delta x^\pm \Delta p^\mp \ge 1
\end{equation}

Light cone coordinates have several useful properties.  A particle with large positive $p_z$ has
large $p^+$ and small $p^-$, and vice-versa for negative $p_z$.  Under Lorentz boosts along the beam axis,
$p^\pm \rightarrow e^{\pm Y_0} p^\pm$, where $\gamma = cosh(Y_0)$.  This means that the rapidity variable
\begin{equation}
  y = {1 \over 2} ln(p^+/p^-)
 \end{equation}
is shifted by a constant under a boost.

An important property of the rapidity follows from the uncertainty principle so that for a wave
$p^\pm \sim 1/x^\mp$.  This means that the space time rapidity is $y \sim \eta$ where 
\begin{equation}
  \eta = {1 \over 2} ln(x^+/x^-)
\end{equation}
This also is true for free streaming particles which originate at the collision point $x_0 = t_0 = 0$,
since for free streaming particles
\begin{equation}
 y = {1 \over 2} ln\{(1+p_z/E)/(1-p_z/E)\} = {1 \over 2} ln\{(1+v_z)/(1-v_z)\} = \eta
\end{equation}
These results have deep implications:  They mean that the momentum space region in which partons form and propagate is strongly correlated with the coordinate space region.  They are in the same region of both
momentum and coordinate space rapidity.

There are also rapidity variables useful for deep inelastic scattering:  Note that $y = ln(p^+/M_T) \sim
ln(\tau_0/x^-)$.  This is useful for thinking about a light cone wavefunction where we would identify
time with $t \sim x^+$ and a time independent wavefunction would have coordinate $x^-$.  Again
the partons are localized in both momentum space and coordinate space rapidity
\begin{figure}[htb]
\begin{center}
  \mbox{{\epsfig{figure=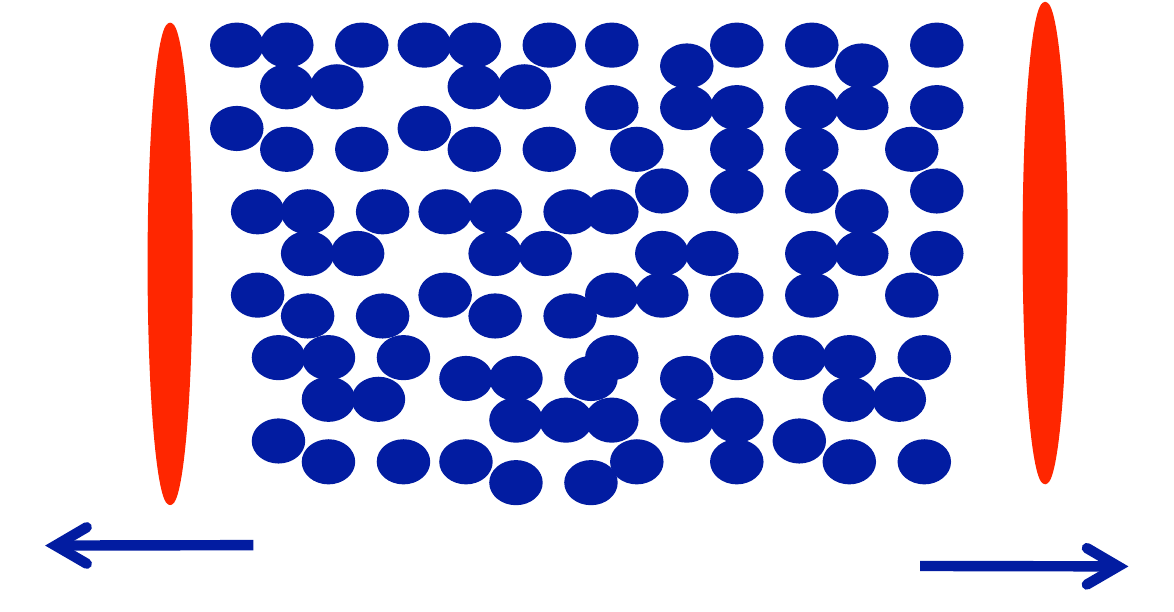, width=0.70\textwidth}}}
   \end{center}
\caption[*]{ \small  The inside-outside cascade.  The fastest particles are formed furthest away from the collision point.  There is an initial momentum gradiant in the system.}
     \label{insideoutside}
\end{figure}

A consequence of the relationship between momentum space rapidity and coordinate space rapidity is that the initial distribution of matter corresponds to an expanding system.  Fast particles are identified with larger values of space time rapidity.  At fixed time, larger space time rapidity corresponds to a position further away from the initial collision point.  The situation is like Hubble expansion in cosmology:  Those particles with highest momentum are furthest away from the collision point.  If it takes some fixed time $\tau_0$ for particles to form, the fastest particles are formed at largest distances from the collision point  This space-time correlation is called an inside-outside cascade.  This initial condition corresponds to an expanding system.  If distributions are independent of rapidity, there is a longitudinally boost invariant expansion of the produced particles.

\section{Implications of a Growing Gluon Density at Small x}

As discussed above, cross sections for gluons rise slowly as the collision energy increases.  The small $x$ gluon distribution rises rapidly.  This means that there must be some very high density of small $x$ gluons in the hadron wavefucntion, and because the density is very large there will be significant self interaction of these gluons.\cite{Gribov:1984tu}-\cite{McLerran:1993ka}  Because the typical separation between the gluons is small, the asymptotic freedom of QCD guarantees that the intrinsic QCD interaction strength will be small,
$\alpha_S << 1$  

The gluon interactions can however be coherent and become quite strong.  An example of an intrinsically weak interaction becoming strong due to coherence is gravity.  The gravitational interaction between two protons is very weak.  The gravitational interaction between you and the earth is quite strong because all the gravitational forces between protons act with the same sign and because the range of the gravitational interaction is very strong.

Suppose we try to add gluons to a proton of some fixed transverse momentum.  We will imagine the DeBroglie wavelength associated with this transverse momentum scale as corresponding to the size of the gluon, $r_0$.  If the gluon was a hard sphere, we might pack the proton with gluons until the density in the transverse plane, $\rho$, was of the order of the inverse area of the gluon $1/r_0^2$.  Gluons however 
are weakly interacting with strength $\alpha_S$, so we will require  a density of order $1/\alpha_S$
in order for the gluons to interact with one another strongly enough so that it is not energetically favorable to add more gluons.   This means that the phase space density of gluons is
\begin{equation}
  {{dN} \over {dyd^2r_T d^2p_T}} \sim {1 \over \alpha_S}
\end{equation}

Quantum mechanically, the phase space density is the occupation number of quantum mechanical states.
Large occupation number means the gluons can be treated classically, as a classical field. The lowest order description is by a classical field theory with a weak coupling.  The strong interactions of the gluon arise because the classical field is strong and has strength $A \sim 1/g$.

\begin{figure}[htb]
\begin{center}
  \mbox{{\epsfig{figure=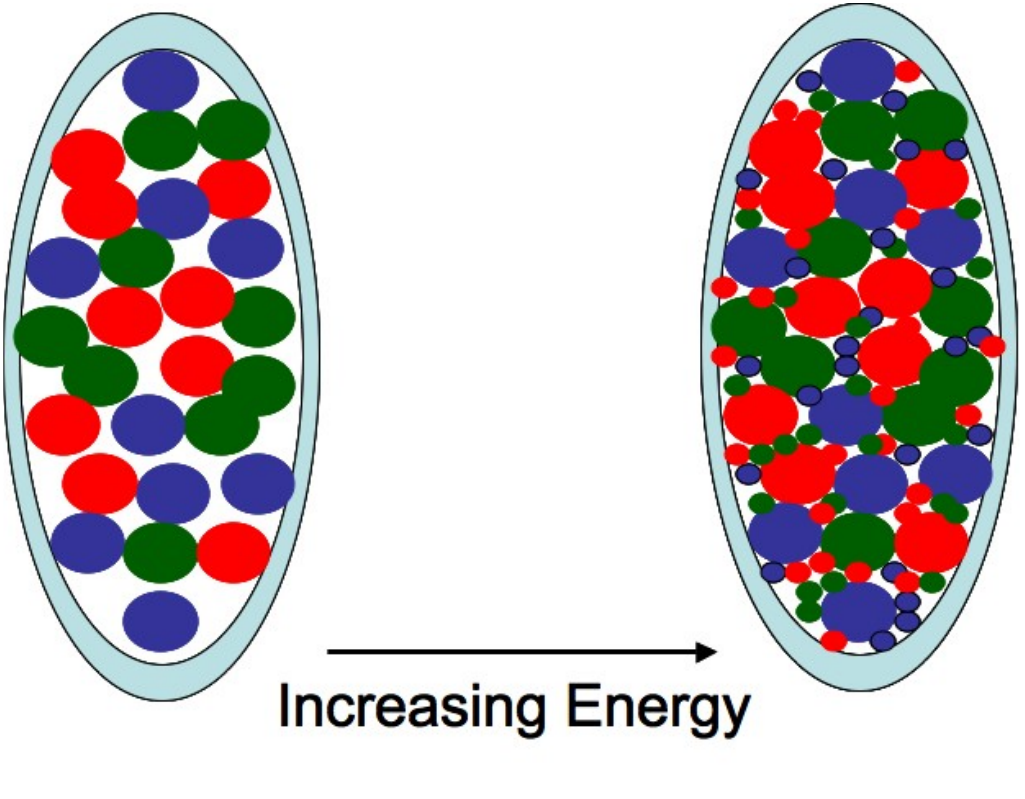, width=0.50\textwidth}}}
   \end{center}
\caption[*]{ \small  Gluon saturation as hadron energy increases. }
     \label{gluesat}
\end{figure}

What happen as we try to add more gluons to the system?  Let us assume all modes with gluon sizes $r > r_0$  are occupied.  This is reasonable because larger size gluons have smaller kinetic energy,
and they should fill the system first.  The new more energetic modes must be smaller and have higher intrinsic $p_T$.  This means that at a fixed hadronic energy, the phase space density is of order $1/\alpha_S$ up to a momentum scale that we call the saturation momentum, $Q_{sat}$.  For modes with higher
momentum, the occupation number is small.  As we increase the hadron energy or alternatively decrease $x$, the saturation momentum should decrease.  The packing of gluons into a hadron is shown in Fig. \ref{gluesat}.

Upon integrating the phase space distribution for the gluons, we find that the total number of saturated gluons is
\begin{equation}
  {1 \over {\pi R^2}} {{dN} \over {dy}} \sim {1 \over \alpha_S} Q_{sat}^2
\end{equation}

\section{Gluon Sources and the Renormalization Group}

The classical gluon fields must arise from sources.  What are these sources?
We imagine we have a gluon field to describe small x degrees of freedom of $x_{min} \le x \le x_1$.
Here $x_{min} \sim \ln(E_{hadron}/\Lambda_{QCD})$, and $x_1$ is some $x$ satisfying $x_1 >> x_0$.
The gluons with larger values of $x$ will not be treated as classical fields but as sources.  Because they have large values of $x$, they can be treated as static sources.  The theory that describes this system must have a distribution of classical sources.  It turns out that this distribution is purely real, and this is a consequence of the time dilation of the sources.

The separation scale $x_0$ is entirely arbitrary.  If we compute quantum corrections to this classical theory, they generate corrections of order $\alpha_S ln(x_1/x_{min})$.  In order that these corrections remain controllable,
they must be small.  The theory we have described is therefore an effective theory, and if we increase the momentum scale, we must somehow decrease $x_1$ so that the corrections remain small, $\alpha_S ln(x_1/x_{min}) << 1$.  The method for changing these scales is the renormalization group, which we shall describe in a later section.\cite{JalilianMarian:1996xn}-\cite{Ferreiro:2001qy}

\section{Color Glass Condensate}

We can now understand the name Color Glass Condensate.  The word Color is because the CGC is composed of colored gluons.  The word Glass is because the classical gluon field is produced by fast moving static sources.  The distribution of these sources is real.  Systems that evolve slowly compared to natural time scales are generically glasses.  A real stochastic distribution of sources that induces an ensemble of classical fields os described by the mathematics of spin glasses.  The word Condensate is because the gluon distribution has maximal phase space density for momentum modes less than the saturation momentum, and the strong gluon fields were self generated by the hadron.

The solution of the renormalization group equation is universal in the small x limit.  This means that the effective theory that describes the Color Glass Condensate is unique, and in that sense the CGC is fundamental.

\section{What Does  a Sheet of CGC Look Like?}

\begin{figure}[htb]
\begin{center}
  \mbox{{\epsfig{figure=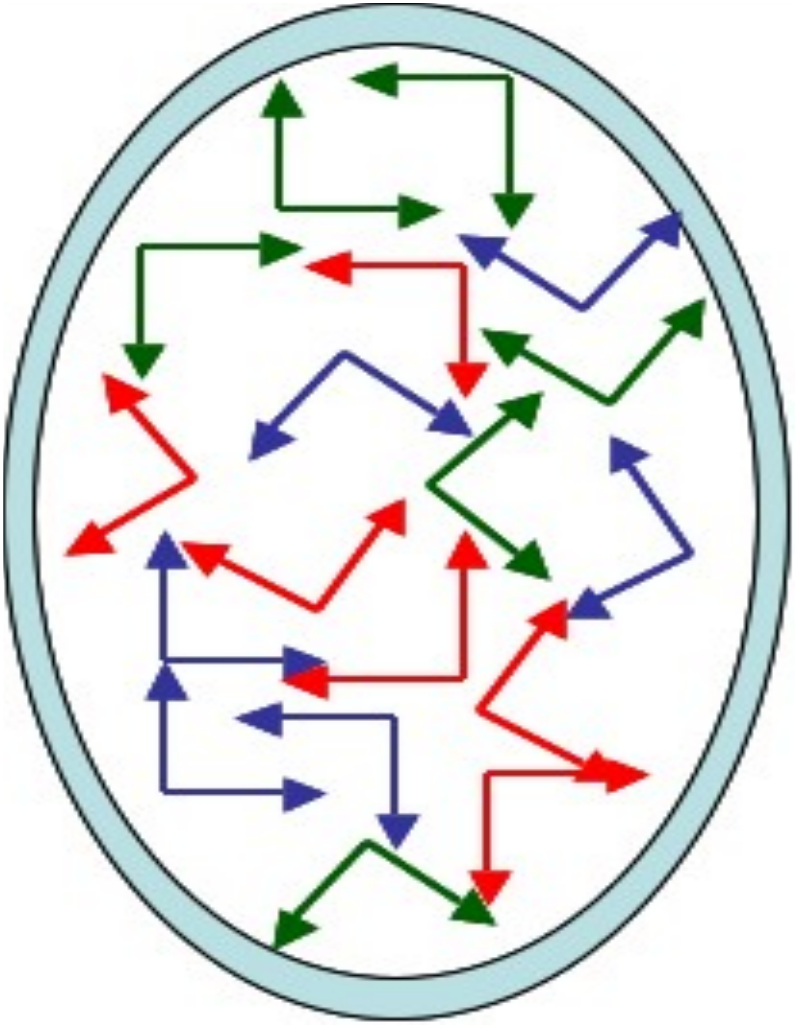, width=0.35\textwidth}}}
   \end{center}
\caption[*]{ \small  A sheet of Colored Glass }
     \label{sheet}
\end{figure}

The sheet on which a CGC sits is Lorentz contracted.  This means that $x^-$ is small.  Onn the other hand,
these fields are static and independent of $x^+$.  This means that the components of 
$F^{i+}$ are big, those of $F^{i-}$ are small and $F^{ij}$ are of order one.  The leading order description 
therefore is when $F^{i+}$ is non-zero and all other fields are approximated as zero.  A little algebra shows
that these fields are of the Lienard-Wiechart form
\begin{equation}
  \vec{E} \perp \vec{B} \perp \hat{z}
\end{equation}
The fields have random polarizations and random colors.  There strength is determined by the distribution of sources, that we shall discuss in a later section.  A sheet of CGC is shown in Fig. \ref{sheet}

\section{The CGC Path Integral}

The correlation functions for the CGC are generated by the path integral
\begin{equation}
 Z = \int_\Lambda~[dA] [d\rho]~ exp\{iS[A,\rho] -F[\rho]\}
\end{equation}
In this equation, the cutoff $\Lambda$ means that the gluon fields are to be considered for longitudinal momentum scales $p^+ \le \Lambda$ corresponding to small x.  The sources on the other hand correspond to larger  $x$ degrees freedom, $ p^+ \ge \Lambda$, that may often be approximated as delta functions in the longitudinal cooredinate $x^-$  These sources generate a current along the light cone of the from
\begin{equation}
  J^\mu = \delta^{\mu +} \rho(x_T, \eta)
\end{equation}
where $\eta$ is the space-time rapidity.

To use this path integral as an effective theory, one simply computes the classical fields associated with the source $\rho$,  inserts the classical field into some quantity one is computing, and then integrates over $\rho$ with the weight function $F$.  One can compute quantum corrections to the classical field, but this will ultimately just change the source distribution $\rho$.  If the corrections are large, then we have used the wrong separation scale $\Lambda$.  The quantum corrections can be reabsorbed into a redefinition of the source distribution function $F$, at the new cutoff scale.

For many purposes, the source distribution function may be treated as a Gaussian
\begin{equation}
 F[\rho] = {1 \over {2\mu}} \rho^2(x,\eta)
 \end{equation}

The renormalization group equation is an equation for
\begin{equation}
   Z_0 = e^{-F[\rho]}
\end{equation}
of the form of a Euclidean time evolution problem
\begin{equation}
 {d \over {d\eta}} = -H[d/d\rho, \rho]~Z_0
\end{equation}
Here $\eta = ln(x^-_0/x^-)$ is a space-time rapidity.

For the strong and intermediate fields strengths that correspond to the CGC, the effective Hamiltonian turns out to be of second order in derivative of $\rho$.  (To derive this Hamiltonian one must do a one loop computation in the presence of arbitrary CGC background fields, and it is beyond the scope of this lecture to present this derivation.)  Unlike an ordinary Hamiltonian, the only terms in the Hamiltonian are a non-linear kinetic energy term, that is two derivatives of $\rho$ multiplied by a non-linear function of $\rho$.
There is no potential functional of $\rho ~$!  Such Hamiltonians describe quantum diffusion.

An example of diffusion is provided by the free Euclidean Schrodinger equation,
\begin{equation}
  {d \over {dt}}~ \psi = - {p^2 \over 2 }~ \psi
\end{equation} 
Because of the lack of a potential, the solution to this equation spreads in time,
\begin{equation}
  \psi \sim e^{-x^2/2t}
\end{equation}
A generic feature of diffusion equations, is that a universal form for the evolving wavefunction is found at large times.  The late time form of the diffusive wavefunction is independent of the initial conditions.
This also means that the  small x gluons never saturate: their wavefunction grows forever as one decreases $x$ corresponding to increased space-time rapidity (corresponding to a Euclidean time)

The evolution equation described above can be applied to compute the rapidity evolution of correlations functions, such as the ones that give the gluon density.  This results in the Balitsky-Kovchegov hierarchy of equations.\cite{Balitsky:1995ub}-\cite{Kovchegov:1999yj}

\section{Phenomenological Implications of the CGC}

There is only space here to briefly outline the implications of the CGC for experiment.  During lectures at this meeting, you will see the theory of the CGC and the Glasma applied to many measured processes so you should expect the detailed descriptions there.

Most important,  the CGC provides a description of the gluon distribution function at small x.  The renormalization group equations derived above allows a computation of the growth of the gluon distribution
functions and for $x\le 10^{-2}$ the description is quite good.  In addition, it describes related processes
such as diffraction.\cite{GolecBiernat:1998js}-\cite{Albacete:2009fh}

The CGC has proved useful in the phenomenology of hadron-hadron collisions.  The saturation momentum scale provides an infrared cutoff for the computation of multi-gluon production.  In naive perturbative QCD, the cross section fro gluon production diverges as $1/p_T^4$ at small $p_T$.  In each hadron, the distribution of soft gluons that is originally $1/p_T$ is tempered to $ln(p_T^2)$.  This is because there are both positive and negative color charges in the hadron, and when measuring at small $p_T$, one is averaging the color charge over large sizes.  In ordinary perturbative QCD, the gluon charges are treated incoherently and there is no cancelation.  This has allowed the computations of multiplicities of particles in heavy ion collisions, and provided initial conditions for the matter produced in such collisions.  Results
are in semi-quantitative accord with experiment.\cite{Kharzeev:2000ph}

There are a variety of phenomena associated with fluctuations and two particle correlations that will be described in other talks at this meeting.  The CGC has provided a computational framework in which such quantities may be evaluated.

\section{The Glasma and Nuclear Collisions}

\begin{figure}[htb]
\begin{center}
  \mbox{{\epsfig{figure=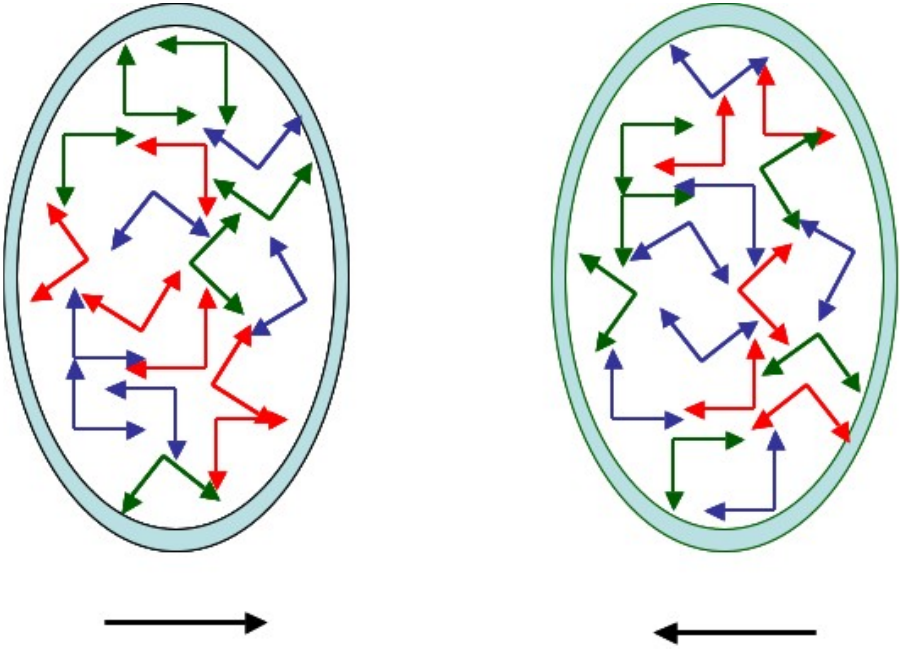, width=0.50\textwidth}}}
   \end{center}
\caption[*]{ \small  The collision of two sheets of Colored Glass. }
     \label{sheetcollision}
\end{figure}

We imagine hadronic collisions as the collision of two sheets of Colored Glass as shown in Fig. \ref{sheetcollision}
What this means is we take the classical field for a hadron propagating along the light cone $z = t$,
and another classical field corresponding to propagating along the light cone $z = -t$ and add them together for the initial classical field.  This solution is valid until the hadrons collide, and then one must
solve the classical Yang-Mills equations in the forward light cone.    I will provide an explicit construction of such solutions in the following paragraphs.\cite{Kovner:1995ja}-\cite{Kovner:1995ts}

We will see that after the hadrons pass through one another, they develop a surface color charge density.
This charge density is equal and opposite on each hadron, and has both color magnetic and color electric charge.  Such charges for both color electric and color magnetic fields must be treated symmetrically since the fields in each hadron involved both fields on an equal footing, and the  Yang-Mills equations are self dual under $E \leftrightarrow B$  
\begin{figure}[htb]
\begin{center}
  \mbox{{\epsfig{figure=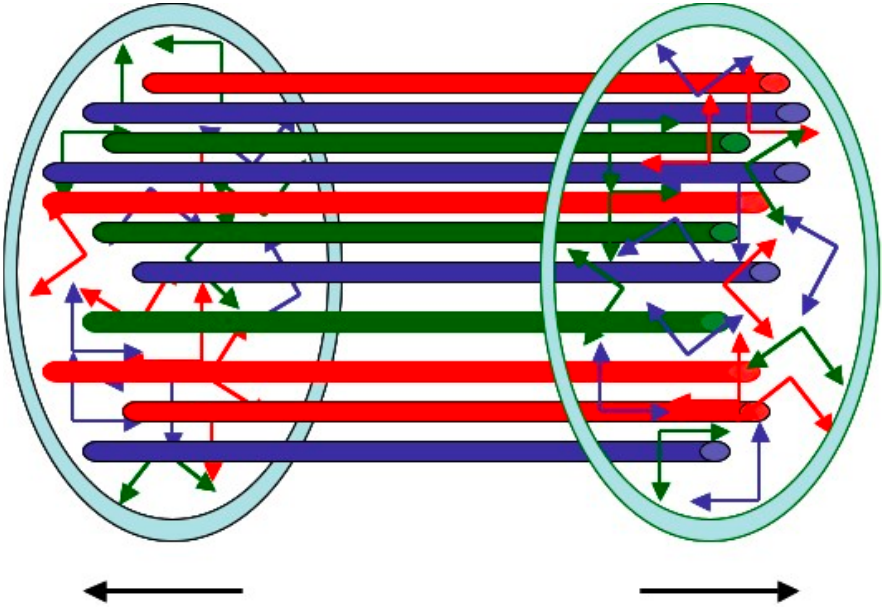, width=0.50\textwidth}}}
   \end{center}
\caption[*]{ \small  The longitudinal color electric and magnetic fields made in hadron collisions. }
     \label{tubes}
\end{figure}

Because the hadrons  have become charged, longitudinal color electric and color magnetic fields will form between them.  This is shown in Fig. \ref{tubes}.  The typical scale of transverse variation of these
fields is the inverse saturation momentum $r \sim 1/Q_{sat}$.  These fields will evolve classically.
They do not need to pair produce particles to decay, because there is both a non-zero $E$
and non-zero $B$ field present.  The Yang-Mills equations are
\begin{equation}
       D^0 \vec{E} = \vec{D} \times \vec{E}
\end{equation}
and $E \leftrightarrow B$.
These fields dilute as the system expands, and when the field strengths are small, the fields may be thought of as quanta of gluons.

The configuration of fields before and after the collision are remarkably different.  Before the collision, the fields are in the sheet of the two Lorentz contracted hadrons.  They are transverse to the collision axis.
After the collision these fields remain, but in addition, there are longitudinal fields.  These fields are produced in the time it takes the sources of color charge $\rho$ to pass through one another.  Recall that the renormalization group arguments above show that such sources sit in rapidity ranges $\mid \Delta  \eta \mid \ge 1/\alpha_S$, so that the time  it takes for the hadrons to pass through one another is $t_{col} \sim e^{-\kappa /\alpha_S}/Q_{sat}$, where $\kappa$ is a constant of order 1. This is a time scale very short compared to the saturation time scale, and in the high energy limit $\alpha_S \rightarrow 0$, may be thought of as infinitesimal.  The formation of these color flux lines is associated with an initial singularity of the high energy limit.  The new matter formed in the collision is prduced from the Color Glass Condensate, has properties remarkably different from the CGC and evolves into the Quark Gluon Plasma.  
It is called the Glasma.\cite{Lappi:2006fp}

\section{Initial Condition for the Glasma}

\begin{figure}[htb]
\begin{center}
  \mbox{{\epsfig{figure=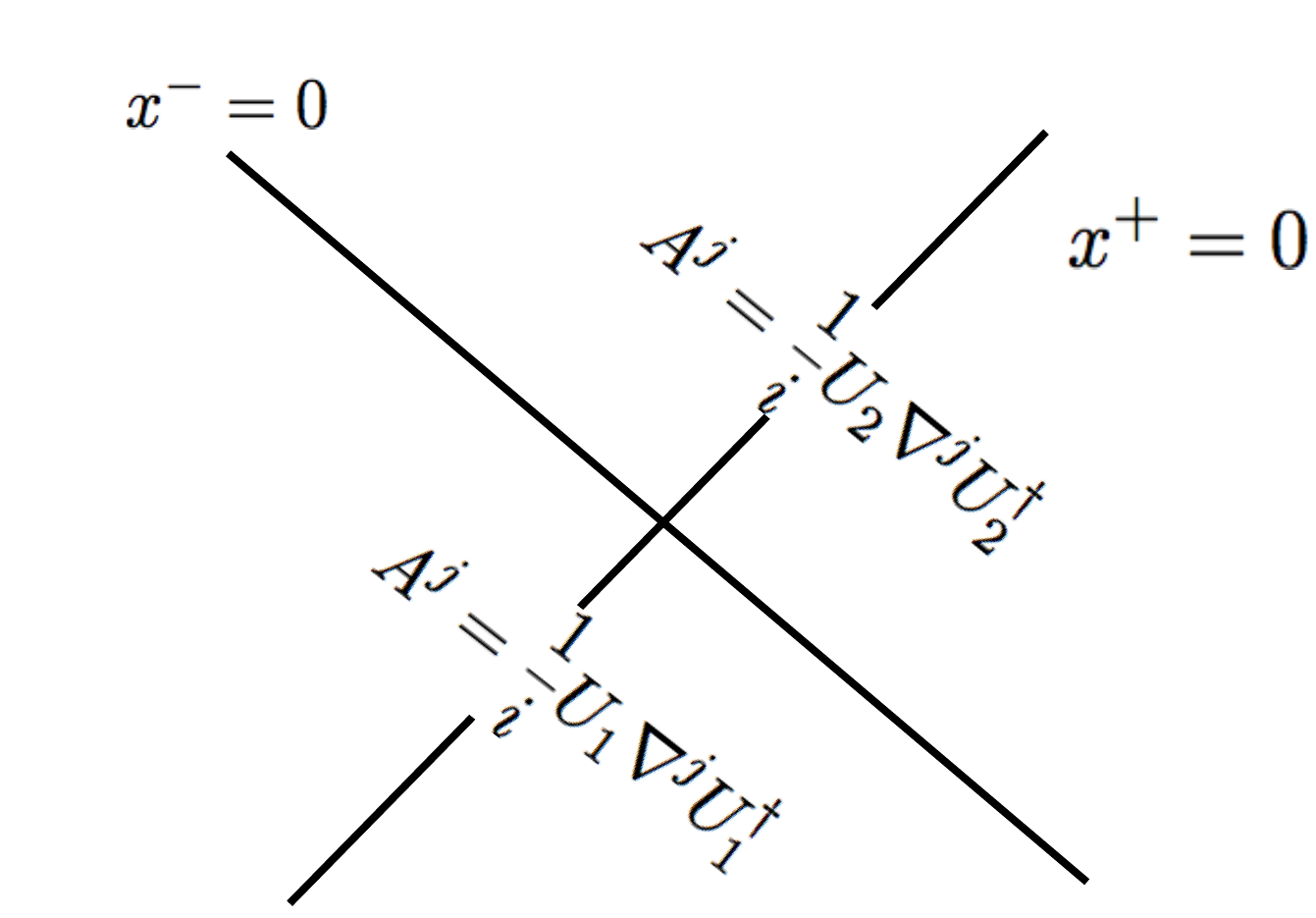, width=0.50\textwidth}}}
   \end{center}
\caption[*]{ \small  The color field for a single hadron. }
     \label{cgcfield}
\end{figure}

Let us recall how we construct a solution for the CGC field corresponding to a single hadron.
We want a solution that has zero color electric and color magnetic field outside the thin sheet on which the color charge density sits.  We will take the charge density to sit at $x^- = 0$.  If we take a purely transverse vector potential that is a pure two dimensional gauge transform of vacuum on either side of 
$x^- = 0$, we will have $F^{\mu \nu} = 0$ outside of $x^- = 0$.  This suggests we look for a solution of the form
\begin{equation}
A^j = \Theta(x^-) {1 \over i} U_2 \nabla^j U^\dagger_2 + \Theta(-x^-) {1 \over i} U_1 \nabla^j U_1 
\end{equation}
As one crosses $x^- =0$, this field has a discontinuity.  This results in a delta function source on the light cone.  (To do this properly, one must spread out the solution in $x^-$.  This is easiest to do for a solution that
is not in light cone gauge, and purely $A^+$, and then gauge transform the result to light cone gauge.)
This construction of the solution is shown in Fig. \ref{cgcfield}

\begin{figure}[htb]
\begin{center}
  \mbox{{\epsfig{figure=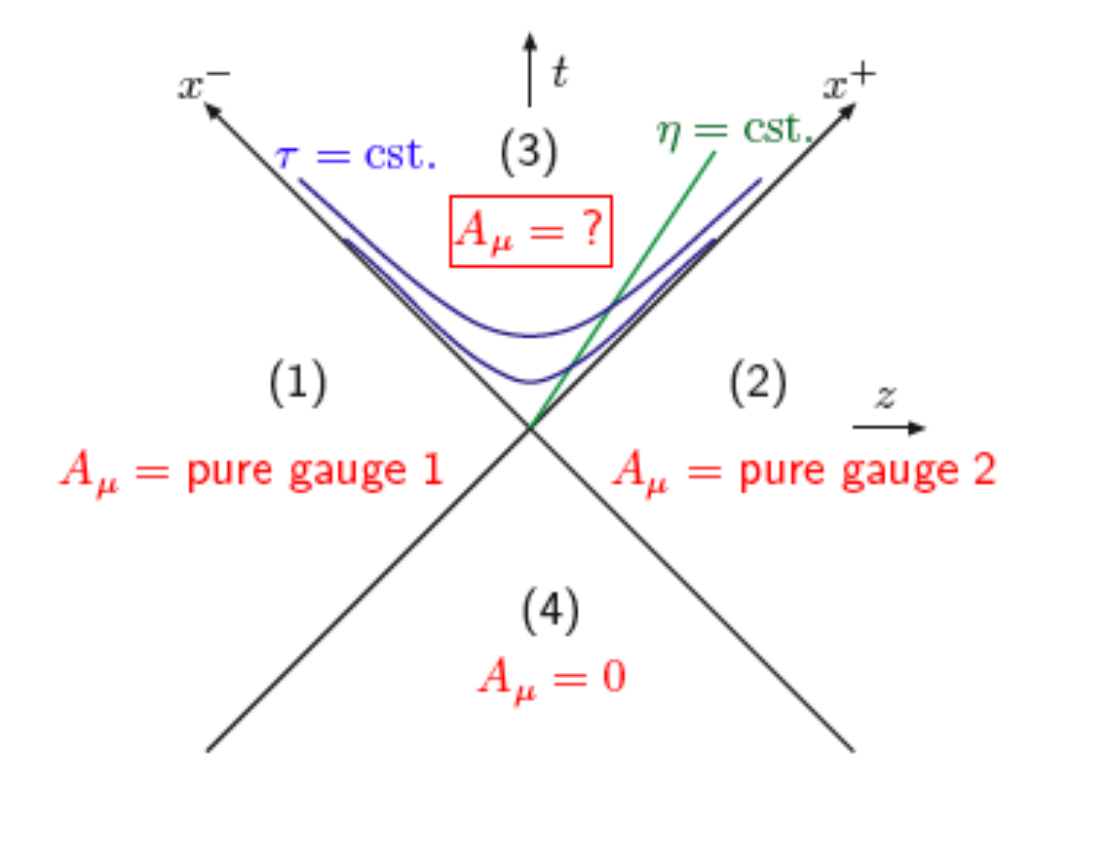, width=0.70\textwidth}}}
   \end{center}
\caption[*]{ \small  The color fields for a two hadrons. }
     \label{glasmafield}
\end{figure}

To construct the color fields for two hadrons, consider Fig. \ref{glasmafield}.  In the backward light cone,
we will take the field to be zero.  This can be done by an overall gauge transformation.  In the side light cones, we take two 2-d gauge transforms of vacuum.  This will generate the correct sources of color charge along the backward light cone.  Unfortunately, we cannot so easily construct the solution in the forward light cone.  Near the edge of the forward light cone, we can try $A^{i} = A^i_1 + A^i_2$.  This indeed will satisfy the equations for the discontinuity to generate the proper charge densities along the forward light cone.  In non-abelian theories, a sum of two gauge transforms of vacuum is not a gauge transform of vacuum, so this will not be a solution in the entire forward light cone.  It provides the initial conditions only for a time evolving solution in the forward light cone.

How do the longitudinal color electric and color magnetic fields arise?
Consider the equation for the electric field
\begin{equation}
  \nabla \cdot E = A \cdot E
\end{equation}
On the backward light cone there are singular terms associated with mutliplying the field times the source, 
but these are always absorbed into the full solution of the equations in the presence of the source.  In the forward light cone there is a new term arising from $A_1 \cdot E_2$ and $A_s \cdot E_2$.  These have delta function pieces because the electric field has a delta function along the sheet.  Therefore there is a surface color charge density induced by the collision.  One can show the density is equal in magnitude but of opposite sign on the two sheets.  In precisely the same manner, longitudinal magnetic fields are induced.

This picture of an initial density of long range color electric and color magnetic fields looks quite similar
to the initial conditions for the Lund Model\cite{Andersson:1983ia}.  There are a few differences.  Here the derivation of the string dynamics comes from weak coupling field theory.  The string hadronizes not by quantum fluctuations but by classical evolution of the Yang-Mills equations.  There are both color electric and color magnetic fields of roughly equal strength. The typical size of the string here is of order $1/Q_{sat}$,
and the typical momentum scale associated with string fragmentation is $Q_{sat}$.

\section{Longitudinal Boost Invariance and Turbulence}

In the forward light cone, a lowest  solution to the classical Yang Mills equation is
\begin{equation}
      A^\pm = x^\pm \alpha_\pm(\tau, x_T)
\end{equation}
and
\begin{equation}
     A^i = \alpha^i(\tau, x_T)
\end{equation}          
These solutions are independent of the rapidity variable $\eta = {1 \over 2} ln(x^+/x^-)$.  They will lead to 
a longitudinally boost invariant description of the matter, and if the matter thermalizes, it would be described by Bjorken boost invariant hydrodynamics.  These equations can be solved for a Guassian fluctuating
initial source density, and might be used as initial conditions for hydrodynamic equations.\cite{Krasnitz:1998ns}-\cite{Lappi:2003bi}  

The solutions to these equations are unstable with respect to the formation of turbulent instabilitiies\cite{Romatschke:2005pm}.  If one inserts a small rapidity dependent perturbation of the boost invariant solution,
it grows rapidly in time.  At some point it becomes larger than the initial Glasma field. Such turbulent instabilities might lead to a fully turbulent system, that might look like a thermally equilibrated system
as far as generating hydrodynamic behaviour.  Not much is yet understand much about the turbulent Glasma nor about its evolution into a thermalized quark gluon plasma.\cite{Dusling:2010rm}

\section{The Glasma and Topological Charge}

The axial vector current is anomalous in QCD,
\begin{equation}
  \partial_\mu J^\mu_5 = {{\alpha_S N_c} \over {4\pi}} F F^d
\end{equation}
where $F^d_{\mu \nu} = {1 \over 2} \epsilon_{\mu \nu \lambda \rho} F^{\lambda \rho}$  This arises from a simple physical consideration:  If there is a nonzero color electric field, a colored particle would be accelerated in the direction of the field.  If there is a B field in the same direction, the particle would rotate around the line of flux.  Therefore chirality is induced.  The same argument implies that chirality is induced 
for antiparticles.  The induced chirality implies a source for the axial vector current.

Such a source of chirality would imply a violation of CP.  Averaged over events and color fields there can of course be no net violation of P and CP.    In fact, one can prove that for the fields of the Glasma, there is no net topological charge.  Such a charge might be generated by fluctuations in initial conditions, that are then amplified by Glasma tubulence.\cite{Kharzeev:2007jp}

One might see the effects of topological charge changing transitions in fluctuations in heavy ion collisions,.  Such fluctuation might be generated by local variation of the topological charge density.
There will be talks in this meeting about the Chiral Magnetic Effect which is one way that one might see such fluctuations.

\section{Phenomenology of the Glasma}

The longitudinal color electric and color magnetic fields may generate two particle correlations in rapidity.
There may also be angular correlations developed by a combination of effects such as hydrodynamic flow,
opacity, or intrinsic angular correlation in the decay of a flux tube.\cite{Dumitru:2008wn}-\cite{Werner:2010aa}  Such angular and rapidity correlations have been seen at RHIC, and there will be much discussion about this in later lectures.  The phenomenon is referred to as the ``ridge" by RHIC experimenters because it appears as a structure long range in rapidity but collimated in azimuthal angle.

The decay of a single flux tube is generates  a negative binomial distribution.\cite{Gelis:2009wh}  The sum of negative binomial distributions is a negative binomial distribution.  Such distributions describe multi-particle production data very well.  The Glasma provide predictions for the parameters of such a distribution.

In addition, the Glasma makes predictions for the multiplicity dependence on energy, and average transverse momenta as a function of energy and multiplicity in high energy collisions.  LHC experimental results will soon test these predictions.  Early data from the LHC concerning  pp collisions show patterns consistent with the CGC hypothesis.  The multiplicity grows with energy, and the transverse momentum scale grows with multiplicity
in a correct way.  In addition, transverse momentum distributions of produced particles approximately scale
when as a function of $p_T/Q_{sat}$.\cite{McLerran:2010ex}

\section*{Acknowledgements}
I gratefully acknowledge my colleagues in Japan who put together this stimulating meeting.
In particular, Yoshi Hatta, Yoshimasa Hidaka, Kazu Itakura, Kenji Fukushima, and Teiji Kunihiro
were very kind and helpful. 
The research of  L. McLerran is supported under DOE Contract No. DE-AC02-98CH10886.

%

\end{document}